\documentstyle[twocolumn,aps,epsfig]{revtex}
\begin{document}
\draft
\title{
Coulomb Gap in the Density of States of Disordered Metals in Two Dimensions}
\author{Peter Kopietz} 
%
\address{
Department of Physics and Astronomy, University of California, Los Angeles,
CA 90095$^{\ast}$\\
Institut f\"{u}r Theoretische Physik, Universit\"{a}t G\"{o}ttingen,
Bunsenstrasse 9, D-37073 G\"{o}ttingen, Germany}
\date{April 21, 1998}
\maketitle
\begin{abstract}
We calculate the effect of Coulomb interactions on the
average density of states $\nu ( \omega )$
of two-dimensional disordered electrons.
It is shown that for weak disorder the most singular terms in the
perturbative expansion of $\nu ( \omega )$
can be summed by means of a simple gauge transformation,
which also establishes a relation between
the low-frequency behavior of $\nu ( \omega )$ and
the average conductivity $\sigma ( \omega )$.
Using this relation, we show 
that if
$\lim_{\omega \rightarrow 0} \sigma ( \omega )
< \infty$,  then 
$\nu ( \omega ) \sim C | \omega | / e^4$
for $\omega \rightarrow 0$, where $C$ is a dimensionless constant and $e$ is the
charge of the electron.
This implies that a normal metallic state
of disordered electrons in two dimensions
is not a Fermi liquid.
\end{abstract}
\narrowtext
%
%
%

In two dimensions arbitrarily weak disorder causes
non-interacting electrons to localize, so that the conductivity vanishes
in the thermodynamic limit\cite{Anderson79}.
For many years it was generally accepted that electron-electron interactions
do not qualitatively change this scenario. The recent experimental
discovery
of a metal-insulator transition in two-dimensional $(2d)$ semiconductor 
devices\cite{Kravchenko95} was therefore a surprise, 
and showed that the current theoretical understanding
of electron-electron interactions in disordered electronic
systems is incomplete. 
The experiment\cite{Kravchenko95} 
was analyzed in the light of a generalized scaling theory\cite{Dobro97},
but a microscopic understanding is still lacking.
Motivated by these exciting new developments, in
this work we shall re-examine the effect of long-range
Coulomb interactions on the average density of states (DOS)
$\nu ( \omega )$ of disordered electrons in two dimensions.

It is well known that Coulomb interactions
drastically modify the DOS in the vicinity of the Fermi energy $\mu$.
For example, in the strongly localized regime 
they give rise to a gap in the DOS at the Fermi energy
(Coulomb-gap),  which 
in $2d$ is of the form\cite{Efros85}
 \begin{equation}
 \nu ( \omega ) \sim  C | \omega | /  e^4 \; \; , \; \; \omega \rightarrow 0
 \; ,
 \label{eq:Cbgap}
 \end{equation}
where $C$ is a dimensionless coefficient and $e$ is the charge of 
the electron\cite{footnote1}.
On the other hand, for weak disorder 
(i.e. for $k_F \ell \gg 1 $, where
$k_F$ is the Fermi wave-vector and $\ell$ is the elastic mean free path)
a perturbative calculation to first order in the dynamically screened
Coulomb interaction yields in two dimensions\cite{Altshuler85},
 \begin{equation}
 \nu ( \omega )  \approx 
 \nu_0 \left[ 1 - \frac{r_0 }{4}
  \ln ( | \omega | \tau_1 )
 \ln ( | \omega | \tau_0 ) 
 \right]
 \; ,
 \label{eq:rhocor}
 \end{equation}
where we have defined
 \begin{equation}
 r_0 = [ (2 \pi )^2 {\cal{D}}_0 \nu_0 ]^{-1} 
 \;  \; ,  \;  \;
 \tau_1 = \tau_0
 [  {\cal{D}}_{0} \kappa^2 \tau_0  ]^{-2}
 \label{eq:r0def}
 \; .
 \end{equation}
Here $\nu_0 = m / 2 \pi $ is average DOS (per spin) at the Fermi energy
of non-interacting electrons with mass $m$,
${\cal{D}}_0 = v_F \ell /2$ is the classical diffusion coefficient
(where $v_F$ is the Fermi velocity),
$\tau_0 = \ell / v_F$ is the elastic lifetime, and
$\kappa$ is the Thomas-Fermi screening wave-vector\cite{footnote2}.
Note that $r_0 = ( \pi k_F \ell )^{-1}$ 
is a dimensionless measure for the resistance of the system
at the frequency scale $\omega \approx \tau_{0}^{-1}$.

Obviously the correction term in Eq.(\ref{eq:rhocor}) 
diverges
for $\omega \rightarrow 0$, indicating the breakdown of perturbation
theory. 
It is tempting to speculate that at sufficiently small frequencies
Eq.(\ref{eq:rhocor}) should eventually cross over
to Eq.(\ref{eq:Cbgap}), at least if
the localization length $\xi$ is finite.
Because for non-interacting electrons in the
weakly localized regime $\xi \propto \exp[ 1 / r_0]$
is exponentially large, 
the crossover should only be
visible at exponentially small frequency scales.
Indeed, resummations of the leading 
logarithms\cite{Finkelstein83,Castellani84,Castellani84b,Belitz93}
in the regime of weak disorder (where $r_0 \ll 1$) 
indicate that in $2d$
the DOS vanishes for $\omega \rightarrow 0$,
but the precise manner in which this happens has 
not been clarified\cite{Belitz93,mistake}.

\begin{figure}
\epsfysize5.3cm 
\hspace{5mm}
\epsfbox{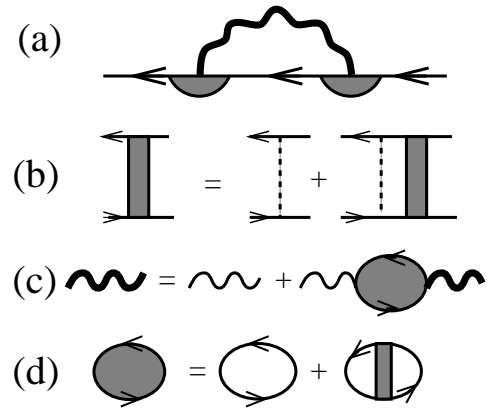}
\vspace{5mm}
\caption{
(a) This exchange correction to the average
Green's function gives rise to the $\ln^2$-correction
to the DOS in Eq.(\ref{eq:rhocor}).
The solid arrows represent non-interacting disorder averaged Green's functions,
the shaded semi-circles denote Diffusion corrections defined
in (b), and the thick wavy line denotes the
dynamically screened interaction defined in (c) and (d).
Here the dashed line denotes impurity scattering, and the thin
wavy line is the bare interaction.
}
\label{fig:diagrams}
\end{figure}

It is instructive to recall the origin of the $\ln^2$-divergence
in Eq.(\ref{eq:rhocor}).
The average DOS of an interacting Fermi system 
can be calculated
from the Fourier transform of the retarded Green's function
$G_{R} ( {\bf{r}} , {\bf{r}}^{\prime} , t )$,
 \begin{equation}
 \nu ( \omega ) = - \frac{1}{\pi} {\rm Im}
\int_{- \infty}^{\infty} d t e^{ i \omega t} \overline{
 G_{R} ( {\bf{r}} , {\bf{r}} , t ) }
 \; ,
 \label{eq:DOS}
 \end{equation}
where the overline denotes averaging over the disorder.
By translational invariance, the average is independent of the position ${\bf{r}}$.
In a perturbative approach,
the corrections to the DOS are obtained
by expanding $G_{R}$ in powers of the interaction
and averaging each term in the expansion over the disorder,
using the impurity diagram technique.
The Feynman diagram responsible for the $\ln^2$-singularity in
Eq.(\ref{eq:rhocor}) is  
shown in Fig.\ref{fig:diagrams}(a).
This  diagram has the special property that at small wave-vectors
($| {\bf{q}} | \ll \ell^{-1})$ and low frequencies $(| \omega | \ll \tau_0^{-1})$
the diffusive motion of the electrons leads to singular corrections to the
charge vertices, which in turn strongly enhance the contribution from
the long-wavelength and low-energy
part of the dynamically screened Coulomb interaction.
The vertex corrections involve so-called Diffuson-diagrams
(Fig.\ref{fig:diagrams}(b)), which
within the Matsubara formalism
renormalize the charge vertices 
for frequencies in the regime $\tilde{\omega}_n ( \tilde{\omega}_n + \omega_m ) < 0$
by a singular factor 
 \begin{equation}
 \Lambda ( {\bf{q}}, i \omega_m ; i \tilde{\omega}_n )
 = 
 [ \tau_0 (
 {\cal{D}}_0 {\bf{q}}^2 + | \omega_m | ) ]^{-1} 
 \label{eq:diffuson}
 \; .
 \end{equation}
Here $\tilde{\omega}_n = 2 \pi ( n + \frac{1}{2} )/ \beta$ 
is the fermionic frequency
carried by the external Green's functions, and $\omega_m = 2 \pi m / \beta$ is the
bosonic frequency carried by the interaction.
$\beta$ is the inverse temperature.
Due to these singular vertex corrections, the main contribution
to the diagram in Fig.\ref{fig:diagrams}(a) comes from energy-momentum
transfers satisfying
 \begin{equation}
 | \omega_m | / {\cal{D}}_0 \kappa \ll | {\bf{q}} | \ll
( | \omega_m | / {\cal{D}}_0 )^{1/2}
\; \; , \; \; | \omega_m | \ll \tau_0^{-1}
 \label{eq:regime}
 \; .
\end{equation}
In this regime, the dynamically screened averaged Coulomb interaction
is within the random-phase approximation (RPA,
see Fig.\ref{fig:diagrams}(c,d))
given by\cite{Altshuler85}
 \begin{equation}
 \overline{f}^{\rm RPA}_{ {\bf{q}} , i \omega_m } \approx (2 {\cal{D}}_0 \nu_0)^{-1} | \omega_m | / 
 {\bf{q}}^2 
 \label{eq:CBsing}
 \; .
 \end{equation}
From Eqs.(\ref{eq:DOS}--\ref{eq:CBsing}) it is 
now straightforward to show
that the exchange diagram in Fig.\ref{fig:diagrams}(a)
indeed gives rise to the $\ln^2$-correction 
in Eq.(\ref{eq:rhocor}).

For our calculation described below 
it is crucial that this
$\ln^2$-correction is entirely due to Coulomb interactions
with momentum transfers small compared with $k_F$. In other words,
the singularity is due to {\it{forward scattering}}.
Recently non-perturbative methods have been developed\cite{Kopietz97}
to study electron-electron interactions with dominant forward scattering
in {\it{clean systems}}.
We shall now show that the functional bosonization approach 
described in Ref.\cite{Kopietz97}(a) can be generalized such that
for weak disorder the dominant corrections to the DOS 
can be summed to all orders in the
dynamically screened Coulomb interaction.

We begin with a number of exact manipulations of the real space,
imaginary time Green's function $G ( {\bf{r}} , {\bf{r}}^{\prime} , \tau - \tau^{\prime} )$
{\it{for a given realization of the disorder}}. 
Representing $G$ as a Grassmannian functional integral,
decoupling 
the interaction in the particle-hole channel
by means of a Hubbard-Stratonovich (HS) field $\phi$, 
and finally integrating over the fermions, we obtain\cite{Kopietz97}(a)
 \begin{equation}
 G ( {\bf{r}} , {\bf{r}}^{\prime} , \tau - \tau^{\prime} )
 = \frac{ \int {\cal{D}} \{ \phi \} e^{ - S \{ \phi \} }
 {\cal{G}} ( {\bf{r}} , {\bf{r}}^{\prime} , \tau , \tau^{\prime} ) }
 { \int {\cal{D}} \{ \phi \} e^{ - S \{ \phi \} }}
 \; ,
 \label{eq:GHS}
 \end{equation}
where the effective action is given by\cite{footnote3}
 \begin{equation}
 S \{ \phi \} = \frac{ {\cal{V}} }{ 2 \beta} \sum_q f_{\bf{q}}^{-1}
 \phi_{-q} \phi_q - {\rm Tr} \ln [ 1 - \hat{G}_0 \hat{V} ]
 \; ,
 \label{eq:Sdef}
 \end{equation}
and
 ${\cal{G}} ( {\bf{r}} , {\bf{r}}^{\prime} , \tau , \tau^{\prime} )$
satisfies
 \begin{eqnarray}
 \left[ - \partial_{\tau} +
 { {\nabla}_{\bf{r}}^2}/{2 m } + \mu - V ( {\bf{r}} , \tau ) -
 U ( {\bf{r}} ) \right]
 {\cal{G}} ( {\bf{r}} , {\bf{r}}^{\prime} , \tau , \tau^{\prime} )
 \nonumber
 \\
 & & \hspace{-40mm}
 = \delta ( {\bf{r}} - {\bf{r}}^{\prime} ) \delta^{\ast} ( {\tau} - {\tau}^{\prime} )
 \label{eq:Gdif}
 \; .
 \end{eqnarray}
Here ${\cal{V}}$ is the volume of the system, 
$f_{\bf{q}} = 2 \pi e^2 / | {\bf{q}} |$ is the Fourier transform
of the bare Coulomb interaction in $2d$, and
$\hat{V}$ and ${\hat{G}}_0$ are infinite matrices in momentum-frequency space.
The matrix elements of $\hat{V}$ 
are $[ \hat{V} ]_{k k^{\prime}} = i \beta^{-1} \phi_{k - k^{\prime}}$, and
$[\hat{G}_{0}]_{k k^{\prime}}$ is the non-interacting
Matsubara Green's function for a given realization of the disorder.
In Eq.(\ref{eq:Gdif}) $U ( {\bf{r}} )$ is the (short-range) disorder
potential,
$V ( {\bf{r}} , \tau ) = i \beta^{-1} \sum_{q} \phi_q$, and
$\delta^{\ast} ( \tau ) = \sum_{n} e^{ i \tilde{\omega}_n \tau }$.

Our strategy is now to separate the {\it{dangerous}} modes of $V ( {\bf{r}} , \tau )$
with energy-momentum transfers in the regime
(\ref{eq:regime}) (which are responsible
for the $\ln^2$-singularity
in Eq.(\ref{eq:rhocor})) from the {\it{harmless}} modes outside this regime. 
Let us therefore define 
$V ( {\bf{r}} , \tau ) = V_{ d} ( {\bf{r}} , \tau ) + 
V_{h} ( {\bf{r}} , \tau )$, where
$V_{d} ( {\bf{r}} , \tau ) = i \beta^{-1} {\sum_{q}}^{\prime} \phi_{q}$,
and the prime indicates that
the sum is restricted to the regime (\ref{eq:regime}).
We then substitute 
into Eq.(\ref{eq:Gdif}) the ansatz
 \begin{equation}
 {\cal{G}} ( {\bf{r}} , {\bf{r}}^{\prime} , \tau , \tau^{\prime} )
 =
 {\cal{G}}_1 ( {\bf{r}} , {\bf{r}}^{\prime} , \tau , \tau^{\prime} )
 e^{ \Phi ( {\bf{r}} , \tau ) - \Phi ( {\bf{r}}^{\prime} , \tau^{\prime} ) }
 \; ,
 \label{eq:schwinger}
 \end{equation}
and seek a solution where $\Phi ( {\bf{r}} , \tau )$ depends only
on the dangerous modes $V_d ( {\bf{r}} , \tau)$. 
It is not difficult to show that one obtains an {\it{exact solution}}
of Eq.(\ref{eq:Gdif}) by choosing ${\cal{G}}_1$ and $\Phi$ such 
that
 $- \partial_{\tau} \Phi ( {\bf{r}} , \tau ) = V_{d} ( {\bf{r}} , \tau )$ and\cite{Kopietz96}
 \begin{eqnarray}
 \left[ - \partial_{\tau} -  
 { ( - i {\nabla}_{\bf{r}} 
 + {\bf{A}} ( {\bf{r}} , \tau ) )^2 }/{2 m } + \mu - V_{h} ( {\bf{r}} , \tau ) -
 U ( {\bf{r}} ) \right]
 \nonumber
 \\
 & & \hspace{-70mm} \times
 {\cal{G}}_1 ( {\bf{r}} , {\bf{r}}^{\prime} , \tau , \tau^{\prime} )
 = \delta ( {\bf{r}} - {\bf{r}}^{\prime} ) \delta^{\ast} ( {\tau} - {\tau}^{\prime} )
 \label{eq:G1dif}
 \; ,
 \end{eqnarray}
where
 ${\bf{A}} ( {\bf{r}} , \tau ) = - i \nabla_{\bf{r}} \Phi ( {\bf{r}} , \tau )$ is a longitudinal
vector potential.
The equation for $\Phi ( {\bf{r}} , \tau )$
can be solved trivially, and we obtain
for the gauge factor in Eq.(\ref{eq:schwinger})
 \begin{eqnarray}
 e^{ \Phi ( {\bf{r}} , \tau ) - \Phi ( {\bf{r}}^{\prime} , \tau^{\prime} ) }
 & = & 
 \nonumber
 \\
 & & \hspace{-25mm}  \exp \left\{
 \frac{i}{\beta} {\sum_{q}}^{\prime} \frac{{\phi}_{q}}{i \omega_m}
 \left[ e^{ i ( {\bf{q}} \cdot {\bf{r}} - \omega_m \tau ) } -
 e^{ i ( {\bf{q}} \cdot {\bf{r}}^{\prime} - \omega_m \tau^{\prime} ) } \right] 
 \right\}
 \label{eq:gaugefactor}
 \; .
 \end{eqnarray} 
Nested exponentials of this type
are familiar from bosonization\cite{Kopietz97}(a).
The crucial observation is now that
Eq.(\ref{eq:gaugefactor}) {\it{contains the leading $\ln^{2}$-singularities to all
orders in perturbation theory}}.
To see this,  note that in Eq.(\ref{eq:G1dif}) we have succeeded
to eliminate from Eq.(\ref{eq:Gdif}) the dangerous part 
$V_{d} ( {\bf{r}} , \tau )$
of the scalar potential
in favour of a longitudinal vector potential. 
Although the associated current vertices can still be dressed by Diffusons, 
the vertex corrections are less singular than in the
case of density vertices\cite{kubo}.
For example, it is not difficult to check that the diagram analogous
to Fig.\ref{fig:diagrams}(a) with current vertices instead of density
vertices does not give rise to any $\ln^2$-singularities.
From Eq.(\ref{eq:schwinger}) it is also obvious that
the gauge factor $e^{ \Phi ( {\bf{r}} , \tau ) -
\Phi ( {\bf{r}}^{\prime} , \tau^{\prime} )}$ cancels
in gauge invariant correlation functions such as the polarization
or the conductivity, so that we immediately see 
that $\ln^2$-singularities
do not appear in the perturbative calculation of
these quantities\cite{Altshuler85,Finkelstein83}.

To make further progress, we have to make several approximations,
{\it{all of which can be systematically improved}}.
(a) First of all, because the
fields
${\bf{A}} ( {\bf{r}} , \tau) $ and $V_{h} ( {\bf{r}} , \tau )$ in Eq.(\ref{eq:G1dif}) do
not generate any $\ln^2$-singularities in perturbation theory, 
let us approximate
${\cal{G}}_1 ( {\bf{r}} , {\bf{r}}^{\prime} , \tau , \tau^{\prime} )
\approx G_{0} ( {\bf{r}} , {\bf{r}}^{\prime} , \tau - \tau^{\prime} )$.
This  is sufficient to  obtain the
leading infrared behavior of the DOS, which is known to be 
dominated by $\ln^{2}$-singularities\cite{Finkelstein83}.
Eq.(\ref{eq:GHS})  can then be written as
$G ( {\bf{r}} , {\bf{r}}^{\prime} , \tau ) = G_{0} ( {\bf{r}} , {\bf{r}}^{\prime} , \tau )
e^{ Q ( {\bf{r}} , {\bf{r}}^{\prime} , \tau ) }$, where
$Q ( {\bf{r}} , {\bf{r}}^{\prime} , \tau ) = \ln \langle 
e^{ \Phi ( {\bf{r}} , \tau ) -
\Phi ( {\bf{r}}^{\prime} , 0 )} \rangle$. Here
$\langle \ldots \rangle$ denotes averaging over the HS-field as defined
in Eq.(\ref{eq:GHS}).
The Debye-Waller factor $Q ( {\bf{r}} , {\bf{r}}^{\prime} , \tau )$ can be calculated systematically
via a linked cluster expansion in powers of the
RPA interaction $f^{\rm RPA}_{ {\bf{q}} {\bf{q}}^{\prime} , i \omega_m }$
{\it{for a given realization of the disorder.}}
For the average DOS we only need
$Q ( 0,0, \tau ) \equiv Q ( \tau )$.
(b) We now calculate $Q ( \tau )$ in Gaussian approximation, retaining only 
the first order term in the linked cluster expansion.
A simple calculation yields
 \begin{equation}
 Q ( \tau ) = - \frac{1}{ \beta {\cal{V}} }
 {\sum_{ {\bf{q}} {\bf{q}}^{\prime} \omega_m  } }^{ \hspace{-1mm} \prime}
 \frac{ f^{\rm RPA}_{ {\bf{q}} {\bf{q}}^{\prime} , i \omega_m}}{ \omega_m^2}
 [ 1 - \cos ( \omega_m \tau ) ]
 \label{eq:Qtaudis}
 \; .
 \end{equation}
In clean systems with dominant forward scattering 
the Gaussian approximation
can be justified via Ward-identities\cite{Kopietz97}.
In the presence of disorder the situation is not so simple.
We shall come back to this problem  below,
where we shall argue
that, at least as long as the 
system has a finite static conductivity, Eq.(\ref{eq:Qtaudis}) 
leads to qualitatively correct results for $\nu ( \omega )$.
Note that by integrating over the HS-field 
{\it{before}} averaging over the disorder we have eliminated the disorder-dependent denominator
in Eq.(\ref{eq:GHS}). 
(c) Finally, we approximately average over the disorder, using 
the factorization $\overline{G ( {\bf{r}} , {\bf{r}} , \tau )}
\approx \overline{ G_0 ( {\bf{r}} , {\bf{r}} , \tau )} 
\exp[  \overline{Q ( \tau )} ]$.
Diagrammatically this amounts to ignoring 
all terms
where impurity lines connect polarization bubbles to other
parts of a given Feynman diagram, a standard approximation
in the usual diagrammatic approach to weakly disordered electrons\cite{Altshuler85}.
Using Eqs.(\ref{eq:r0def}) and (\ref{eq:CBsing}),
we obtain 
in the limit ${\cal{V}} \rightarrow \infty$ and $\beta \rightarrow \infty$
 \begin{equation}
 \overline{Q ( \tau )} =  \frac{1}{2} \int_{0}^{\tau_0^{-1}} \frac{ d \omega}{\omega}  r_0 \ln ( 
 \omega / {\cal{D}}_0 \kappa^2   ) \left[ 1 - \cos ( \omega \tau ) \right]
 \; .
 \label{eq:Qtau}
\end{equation}
For large $\tau$ this implies
$ \overline{Q ( \tau )} \sim - \frac{r_0}{4} \ln ( \tau / \tau_1)
 \ln ( \tau / \tau_0 )
$,
where $r_0$ and $\tau_1$ are defined in Eq.(\ref{eq:r0def}).
From Eq.(\ref{eq:DOS}) we finally obtain
after analytic continuation 
 \begin{equation}
 \nu ( \omega )  \approx  \nu_0 \frac{2}{\pi} \int_{\tau_0}^{\infty} 
 dt \frac{ \sin ( | \omega | t )}{t} 
\exp \left[
 - \frac{r_0}{4} \ln ( t / \tau_1) \ln ( t / \tau_0 )
    \right]
 \label{eq:disfinal}
 \; .
 \end{equation}
Expanding 
the right-hand side of Eq.(\ref{eq:disfinal}) 
to first order in $r_0$, we recover
the perturbative result (\ref{eq:rhocor}),
which is valid as long as
$- \ln ( | \omega | \tau_0 )  
{ \raisebox{-0.5ex}{$\; \stackrel{<}{\sim} \;$}} 
\sqrt{ 1/ {r}_0 }$.
Keeping in mind that $1/r_0 \gg 1$, we find that there exists an
intermediate frequency range
$\sqrt{ 1/ r_{0} } 
{ \raisebox{-0.5ex}{$\; \stackrel{<}{\sim} \;$}} 
- \ln ( | \omega | \tau_0 ) 
{ \raisebox{-0.5ex}{$\; \stackrel{<}{\sim} \;$}} 
1/ r_0$ where
Eq.(\ref{eq:disfinal}) can be approximated by 
 \begin{equation}
 \nu ( \omega ) \approx  \nu_0 \exp \left[ - \frac{r_0}{4} 
\ln ( | \omega | \tau_1 ) \ln ( | \omega | \tau_0 ) \right]
 \; ,
 \label{eq:nuint}
 \end{equation}
in agreement with Finkelstein\cite{Finkelstein83}.
This is a non-trivial check that 
the gauge factor (\ref{eq:gaugefactor})
indeed contains the dominant singularities.
Note that we have derived Eq.(\ref{eq:nuint}) without using replicas. 
For frequencies below the exponentially small scale
$\tau_0^{-1} \exp[  - 1 / r_0 ]$
Eq.(\ref{eq:nuint}) is not valid\cite{Finkelstein83,Belitz93}.
It is easy to show, however, that in this regime Eq.(\ref{eq:disfinal})  
smoothly crosses over to the
linear Coulomb gap given in Eq.(\ref{eq:Cbgap}), with
the numerical constant 
$C = 4 \pi^{-1/2} \sqrt{ r_0} \exp[  {1/r_0} ]$.
The amazing fact is that for $\omega \rightarrow 0$ 
the prefactor $\nu_0$ in Eq.(\ref{eq:disfinal}) 
disappears and is replaced by
$| \omega | / e^4$, which in $2d$ has the same
units as $\nu_0$.

We now argue that
{\it{if the conductivity $\sigma ( \omega )$ does not 
diverge for $\omega \rightarrow 0$}}, then
the true asymptotic low-frequency behavior  of the DOS
is indeed given by Eq.(\ref{eq:Cbgap}),
although our above result for the numerical value of $C$ is not reliable.
Let us therefore recall that the parameter $r_0$
in Eq.(\ref{eq:Qtau}) is the
dimensionless resistance {\it{at frequency scale $\tau_0^{-1}$}}.
With the above approximations (b) and (c)
we have effectively replaced 
$S \{ \phi \}$ in Eq.(\ref{eq:GHS}) by
its  Gaussian average, which in the regime (\ref{eq:regime}) can be written as
 \begin{equation}
 \overline{S} \{ \phi \} \approx \frac{ {\cal{V}} }{ \beta (2 \pi)^2  }
 {\sum_q}^{\prime}  \frac{ {\bf{q}}^2 }{ r_0 | \omega_m | } 
 \phi_{-q} \phi_q
 \; .
 \label{eq:Seffav}
 \end{equation}
In deriving Eq.(\ref{eq:Qtau}), we have assumed
that $r_0$ is not renormalized for $\omega \rightarrow 0$,
which is in general incorrect:
higher order corrections in the disorder, as well as the
non-Gaussian terms neglected in Eq.(\ref{eq:Seffav})
renormalize the effective value of
$r_0$ at small frequencies,
so that
in Eq.(\ref{eq:Seffav}) we should replace 
$r_0 \rightarrow r ( i \omega_m  )$. 
Note that $r^{-1} ( \omega )$ is proportional to the
frequency-dependent conductivity. A microscopic calculation
of $r ( \omega  )$ is beyond
the scope of the methods developed in this work.
However, we may assume different scenarios for $r ( \omega )$ and
calculate the consequences for the DOS.

First of all, suppose that
$0 < \lim_{\omega \rightarrow 0} r ( \omega  ) \equiv r_{\ast} < \infty$,
so that the system is a
{\it{normal metal}}, with a finite static conductivity\cite{Castellani98}.
In this case
the non-Gaussian terms neglected in Eq.(\ref{eq:Seffav}) are irrelevant
(in the renormalization group sense) with respect to the Gaussian fixed point
action\cite{Kane91}. 
We conclude that
{\it{if}} interactions stabilize a normal metallic state in $2d$\cite{Castellani98},
{\it{then}}
the DOS exhibits a Coulomb gap 
$\nu ( \omega ) \sim C | \omega | / e^4$, just like
in the strongly localized regime.
The numerical value of $C$ depends
on $r_{\ast}$, which we have not calculated.
Because the DOS vanishes at the Fermi energy,
a normal metal in $2d$ (if it exists) is not
a conventional Fermi liquid. 
Let us emphasize that 
for our calculation
we have assumed that the screening length
$\kappa^{-1}$ is smaller than the mean free path\cite{footnote2}.
In contrast, in the
strongly localized regime the screening length is infinite
and the Coulomb gap
can be explained from the 
classical Hartree energy\cite{Efros85}.
Thus, the physics responsible for the Coulomb gap is different
in both cases.
Numerical evidence that in $2d$ the Coulomb gap
in the DOS
survives in a non-Fermi liquid metallic state has
been found previously by Efros and Pikus\cite{Efros95}.

Depending on the low-frequency behavior of $r ( \omega )$, there are
two other possibilities.
If $r ( \omega ) \rightarrow 0$ for $\omega \rightarrow 0$, the system
is a {\it{perfect metal}}. 
This is the physically most plausible scenario, because
it is consistent with the
generalized scaling theory\cite{Dobro97}. 
In $2d$ we expect
that the divergence
of $\sigma ( \omega )$ is logarithmic, so that
$r ( \omega ) \sim - \gamma /   \ln ( \omega \tau_0)$ with
$\gamma > 0$.
Because the weak logarithmic singularity does not affect
the irrelevance of the non-Gaussian terms neglected in
Eq.(\ref{eq:Seffav}), we may calculate 
$\nu ( \omega )$ by substituting
$r_0 \rightarrow - \gamma / \ln  ( \omega \tau_0)$
in Eq.(\ref{eq:Qtau}).
This yields
$\nu ( \omega ) \propto | \omega |^{\gamma / 2}$.
Note that in Ref.\cite{Finkelstein83} Finkelstein
found (incorrectly\cite{Castellani84b,mistake}) 
$r ( \omega ) \sim - \frac{1}{2} / \ln ( \omega \tau_0 )$.
In this case we obtain $\nu ( \omega ) \propto | \omega |^{1/4}$, in
agreement with Ref.\cite{Finkelstein83}.
The third possibility is
$r ( \omega ) \rightarrow \infty$ for $\omega \rightarrow 0$, so that
the system is an {\it{insulator}}.
Because we know that
Eq.(\ref{eq:Cbgap}) is valid
for strongly localized electrons \cite{Efros85}, {\it{and
remains correct even if $\sigma ( 0 )$ is finite}},
by continuity it is extremely plausible that 
$\nu ( \omega ) \sim C | {\omega } | / e^4$ 
in the entire localized regime.

Experimentally $\nu ( \omega )$ can be obtained
by measuring the tunneling conductance $dI/dV$ as function
of the applied voltage $V$.
Such a measurement was performed in Ref.\cite{White85}, 
and the measured $dI / dV$ was compared with 
the perturbative prediction (\ref{eq:rhocor}).
Keeping in mind that a logarithmic 
dependence was only observed in a rather small
voltage interval, it seems  that
the data
shown in Fig.1 (a) of Ref.\cite{White85} are consistent with 
the emergence of a linear Coulomb gap (\ref{eq:Cbgap}).
It would be very interesting to measure the tunneling conductance
in the $2d$ materials that exhibit a metal-insulator transition\cite{Kravchenko95}.
On the metallic side of the transition, a linear dependence
of $dI/dV$ on the voltage $V$
would be consistent with a finite value of the static conductivity,
while a non-linear dependence
would be consistent with $\sigma ( 0 ) = \infty$.

I would like to thank S. Chakravarty for many discussions 
and for pressing me to sharpen my arguments.
I am also thankful to 
A. H. Castro-Neto and G. E. Castilla for their hospitality
during a visit in Riverside, where part of this work
was carried out,
and to S. Kivelson for
drawing my attention to the work quoted in Ref.\cite{Kane91}. 
This work was financially supported by the DFG.

%

%
%
%

\begin{thebibliography}{99}
%
\bibitem[*]{address}
Address from October 1997 - April 1998.
%
\bibitem{Anderson79}
E. Abrahams {\it{et al.}},
Phys. Rev. Lett. {\bf{42}}, 673 (1979).
%
\bibitem{Kravchenko95}
S. V. Kravchenko {\it{et al.}}, Phys. Rev. B {\bf{51}}, 7038 (1995).
%
\bibitem{Dobro97}
V. Dobrosavljevi\'{c} {\it{et al.}},
Phys. Rev. Lett. {\bf{79}},
455 (1997).
%
\bibitem{Efros85}
A. L. Efros and B. I. Shklovskii,
in {\it{Electron-Electron Interactions in Disordered Systems}},
edited by A. L. Efros and M. Pollak (North-Holland, Amsterdam, 1985).
%
\bibitem{footnote1}
We measure energies relative to $\mu$ and set
$\hbar = 1$.
%
\bibitem{Altshuler85}
B. L. Altshuler and A. G. Aronov, in Ref.\cite{Efros85}.
%
\bibitem{footnote2}
For our calculation we assume
$r_0 \ll 1$ and
$\ell^{-1} \ll \kappa \ll k_F$, so that
$ {\cal{D}}_0 \kappa^2 \tau_0 = ( \kappa \ell )^2/2 \gg 1$
and $\tau_1 \ll \tau_0$.
In two dimensions $\kappa = 2 \pi e^2 \partial n / \partial \mu$, where
$ \partial n / \partial \mu$  is the compressibility. 
Note that for non-interacting electrons $\partial n / \partial \mu = 2 \nu_0$, but
in general there is no simple relation 
between $\partial n / \partial \mu$ and the DOS\cite{Finkelstein83}.
%
\bibitem{Finkelstein83}
A. M. Finkelstein, Zh. Eksp. Teor. Fiz. {\bf{84}}, 168 (1983)
[Sov. Phys. JETP {\bf{57}}, 97 (1983)].
%
\bibitem{Castellani84}
C. Castellani {\it{et al.}},
Phys. Rev. B {\bf{30}}, 527 (1984).
%
\bibitem{Castellani84b}
C. Castellani {\it{et al.}}, Phys. Rev. B {\bf{30}}, 1596 (1984);
A. M. Finkelstein, Z. Phys. B {\bf{56}}, 189 (1984).
%
\bibitem{Belitz93}
D. Belitz and T. R. Kirkpatrick, Phys. Rev. B {\bf{48}}, 14072 (1993).
Note that in the exponent of Eq.(4.12) of this work 
one should replace $1/2 \rightarrow {1}/{32}$. I would like to thank
D. Belitz and T. R. Kirkpatrick for checking their notes and confirming this printing error.
%
\bibitem{mistake}
Finkelstein\cite{Finkelstein83} found 
$\nu ( \omega ) \propto | \omega |^{1/4}$ for a model where
the back-scattering channel is suppressed.
In obtaining this result he used the fact that
the conductivity $\sigma ( \omega )$ seemed to diverge for $\omega \rightarrow 0$.
Subsequently a mistake in the renormalization group
equations of Ref.\cite{Finkelstein83} was discovered\cite{Castellani84b},
and it was shown that the corrected equations actually
imply that $\lim_{\omega \rightarrow 0} \sigma ( \omega )$ is finite. 
%
\bibitem{Kopietz97}
For recent reviews see (a) P. Kopietz,
{\it{Bosonization of Interacting Fermions in Arbitrary
Dimensions}}, (Springer, Berlin, 1997), and
(b) W. Metzner, C. Castellani, and 
C. Di Castro, Adv. Phys. {\bf{47}}, 317 (1998).
%
\bibitem{footnote3}
We use collective labels
$q = [ {\bf{q}} , i \omega_m]$ and $k = [ {\bf{k}} , i \tilde{\omega}_n ]$.
%
\bibitem{Kopietz96}
In 
P. Kopietz and G. E. Castilla, Phys. Rev. Lett. {\bf{76}}, 4777 (1996)
(see also Ref.\cite{Kopietz97}(a))
the ansatz (\ref{eq:schwinger}) was used to 
include the effect of the curvature of the Fermi surface
in higher-dimensional bosonization.
In this case a different choice of $\Phi ( {\bf{r}} , \tau )$ was useful.
%
\bibitem{kubo}
This is also the reason why in the calculation of $\sigma ( \omega )$ 
via the Kubo formula it is convenient to
choose a gauge where the electric field is represented
as a vector potential.
%
\bibitem{Castellani98}
C. Castellani, C. Di Castro, and P. A. Lee, Phys.
Rev. B {\bf{57}}, R9381 (1998). 
%
\bibitem{Kane91}
To see this, let us approximate the $n^{\rm th}$-order 
vertices in the expansion of $\overline{S} \{ \phi \}$ by constants
$u_n$ and calculate their scaling dimensions.
Keeping the Gaussian part invariant under
$ {\bf{q}}^{\prime} = b {\bf{q}}$,
$ {\omega}^{\prime}_m = b^{z} \omega_m$ 
(where $b > 1$ is the length rescaling factor and $z=2$ is the dynamic exponent),
it is easy to see that $u_n^{\prime} = b^{- 2(n-2)} u_n$.
Thus, the non-Gaussian terms ($n > 2$) decrease as we lower the energy and momentum scale,
so that the Gaussian action correctly describes the
infrared physics. See
C. L. Kane {\it{et al.}},  Phys. Rev. B {\bf{43}}, 3255 (1991)
for a similar analysis. 
%
%
%
\bibitem{Efros95}
A. L. Efros and F. G. Pikus, Solid State Commun. {\bf{96}}, 183
(1995). I would like to thank A. L. Efros for
pointing this work out to me.
%
\bibitem{White85}
A. E. White, R. C. Dynes, and J. P. Garno,
Phys. Rev. B {\bf{31}}, 1174 (1985).
%
%
\end{thebibliography}
\end{document}